\documentclass[11pt]{article}

\usepackage{authblk}

\usepackage[top=50mm, bottom=50mm, left=50mm, right=50mm]{geometry}
\usepackage{lineno}
\usepackage{amssymb}
\usepackage{amsmath}
\usepackage{amsthm}
\usepackage{epsfig}
\usepackage{graphicx}
\usepackage{graphics}
\usepackage{float}
\usepackage{multirow}
\usepackage{color}
\usepackage{lineno}
\usepackage{fullpage}
\usepackage[normalem]{ulem} 
\usepackage{makeidx}
\usepackage{xspace}

\usepackage{wrapfig}

\usepackage[colorlinks=true, linktoc=page, linkcolor=blue, citecolor=blue, urlcolor=blue]{hyperref}

\def\GeV{{\text{ }\mathrm{GeV}}}
\def\MeV{{\text{ }\mathrm{MeV}}}
\def\keV{{\text{ }\mathrm{keV}}}

\newcommand{\epem}{e^+e^-}
\newcommand{\alphas}{\alpha_{S}}

\newcommand{\alphasmZ}{\alphas(m_\mathrm{Z})}
\newcommand{\lqcd}{\Lambda_{_\mathrm{QCD}}}

\newcommand{\sqrts}{\sqrt\mathrm{s}}

\newcommand{\MW}{m_\mathrm{W}}
\newcommand{\MZ}{m_\mathrm{Z}}

\newcommand{\qqbar}{\ensuremath{q\overline{q}}}

\newcommand*{\eg}{e.g.\@\xspace}
\newcommand*{\ie}{i.e.\@\xspace}

\newcommand{\gfitter}{\textsc{gfitter}}
\newcommand{\mathematica}{\textsc{Mathematica}}
\newcommand{\nin}{\noindent}

\newcommand{\Vcs}{|V_{cs}|}
\newcommand{\Vcd}{|V_{cd}|}
\newcommand{\Rlz}{\mathrm{R}_\mathrm{Z}}
\newcommand{\RZexp}{\mathrm{R}_\mathrm{Z}^\mathrm{exp}}
\newcommand{\RW}{\mathrm{R}_\mathrm{W}}
\newcommand{\so}{\sigma_\mathrm{Z}^\mathrm{had}}
\newcommand{\GZ}{\Gamma_\mathrm{Z}^\mathrm{tot}}

\newcommand{\GZhad}{\Gamma_\mathrm{Z}^\mathrm{had}}

\providecommand{\GWhad}{\Gamma_\mathrm{W}^\mathrm{had}}
\newcommand{\GWtot}{\Gamma_\mathrm{W}^\mathrm{tot}}

\def\cO#1{{{\cal{O}}}\left(#1\right)}
\normalbaselines

\usepackage{caption} 
\begin{document}

\title{Improved strong coupling determinations from hadronic\\ 
decays of electroweak bosons at N$^3$LO accuracy}

\author[1]{David~d'Enterria\footnote{Corresponding author: David.d'Enterria@cern.ch}}
\author[2]{Villads Jacobsen}
\affil[1]{CERN, EP Department, CH-1211 Geneva, Switzerland}
\affil[2]{Institute for Physics \& Astronomy, Aarhus University, DK-8000 Aarhus, Denmark}

\maketitle

\setcounter{page}{1}

\maketitle

\begin{abstract}
\nin We present two new extractions of the QCD coupling constant at the Z pole, $\alphasmZ$, from detailed comparisons of inclusive W and Z hadronic decays data to state-of-the-art perturbative Quantum Chromodynamics calculations at next-to-next-to-next-to-leading order (N$^{3}$LO) accuracy, incorporating the latest experimental and theoretical developments. In the W boson case, the total width computed at N$^{3}$LO is used for the first time in the extraction. For the Z boson pseudo-observables, the N$^{3}$LO results are complemented with the full two- and partial three-loop electroweak corrections recently made available, and the experimental values are updated to account for newly estimated LEP luminosity biases. A combined reanalysis of the Z boson data yields $\alphasmZ = 0.1203 \pm 0.0028$, with a 2.3\% uncertainty reduced by about 7\% compared to the previous state-of-the-art. From the combined W boson data, a value of $\alphasmZ = 0.101 \pm 0.027$ is extracted, with still large experimental uncertainties but also reduced compared to previous works. The levels of theoretical and parametric precision required in the context of QCD coupling determinations with permil uncertainties from high-statistics W and Z boson samples expected at future $\epem$ colliders such as the FCC-ee, are discussed in detail.
\end{abstract}

\section{Introduction}

The coupling constant $\alphas(Q)$ determines the strength of the strong interaction between quark and gluons, described theoretically by Quantum Chromodynamics (QCD), at a given energy scale $Q$. The energy dependence, or ``running'', of $\alphas(Q)$ can be precisely predicted from the QCD $\beta$ function evolution, which is theoretically known up to five-loops accuracy today~\cite{Baikov:2016tgj,Luthe:2016ima,Herzog:2017ohr}.
The QCD coupling evaluated commonly at the reference energy scale of the Z mass, $\alphasmZ$, is one of the fundamental parameters of the Standard Model (SM). Its value not only chiefly impacts the theoretical calculations of all scattering and decay processes involving real and/or virtual quarks and gluons~\cite{alphas_confs}, but it also plays a role \eg\ in the stability of the electroweak vacuum~\cite{Buttazzo:2013uya}. 
Known today with a 0.9\% precision, $\alphasmZ = 0.1179~\pm~0.0010$, the QCD coupling is the worst known of all fundamental interaction couplings in nature~\cite{PDG}, and such an imprecision propagates as an input parametric uncertainty in the calculation of many important physics observables, in particular in the electroweak (EW), Higgs, and top-quark sectors of the SM~\cite{Blondel:2019vdq,Blondel:2019qlh}. The current world-average $\alphasmZ$ is derived~\cite{PDG} from a combination of seven subclasses of approximately independent observables measured in (i) $\epem$ collisions (hadronic Z boson and $\tau$ decays, plus event shapes and jet rates), (ii) electron-proton ($e$p) deep-inelastic scattering DIS (structure functions, and global fits of parton distributions functions PDFs),  (iii) proton-proton (p-p) collisions (inclusive top-pair cross sections), (iv) heavy quarkonia decays, as well as from (v) lattice QCD computations constrained by the empirical values of hadron masses and decay constants. In order to be combined into the $\alphasmZ$ world average, the experimental and lattice results need to have a counterpart perturbative QCD (pQCD) prediction with, at least, a next-to-next-to-leading-order (NNLO) accuracy.\\

Among the $\alphas$ extraction methods, those based on inclusive hadronic decays of the electroweak bosons are arguably the ``cleanest'' ones from both theoretical and experimental perspectives. This is so because: (i) different observables related to hadronic W and Z decays can be very accurately measured in high-energy $\epem$ collisions (provided one has large enough data samples), and (ii) the corresponding theoretical predictions can be computed with a very high theoretical accuracy (beyond NNLO today) with suppressed non-pQCD effects thanks to the large energy scale given by the electroweak masses ($m_\mathrm{W,Z}\gg \lqcd\approx 0.2$~GeV). The following high-precision electroweak boson observables are commonly used to extract $\alphasmZ$ from the corresponding data-theory comparisons~\cite{alphas_confs,PDG}: 
\begin{itemize}
\item The W and Z hadronic widths, theoretically computable via the generic expression
\begin{eqnarray}
\Gamma^\mathrm{had}_\mathrm{W,Z}(Q) = 
 \Gamma^{^\mathrm{Born}}_\mathrm{W,Z} \left(\! 1 +\sum^{4}_{i=1} a_i(Q)\left(\!\frac{\alphas(Q)}{\pi}\right)^i\!\!+ \mathcal{O}(\alphas^5)+\delta_{_\mathrm{EW}}+\delta_\mathrm{mix}+\delta_\mathrm{np}\right)\!,
\label{eq:Gamma_alphas}
\end{eqnarray}
where the Born width $\Gamma^{^\mathrm{Born}}_\mathrm{W,Z}=f(G_F,\,m_\mathrm{W,Z}^3,\,N_C;\mathrm{\sum |V_{ij}|^2})$ depends on the Fermi constant $G_F$, (the cube of the) EW boson masses, the number of colours $N_C$, and, in the W case, also on the sum of CKM matrix elements $\rm|V_{ij}|^2$. The $a_i(Q)$ and $\delta_\mathrm{EW,mix,np}$ terms are, respectively, higher-order QCD, EW, mixed, and non-pQCD corrections discussed below. 

Since the total W and Z widths ---given by the sum of hadronic and leptonic partial widths $\Gamma^\mathrm{tot}_\mathrm{W,Z} = \Gamma^\mathrm{had}_\mathrm{W,Z} + \Gamma^\mathrm{lep}_\mathrm{W,Z}$--- have smaller experimental uncertainties than the hadronic one alone, and since $\Gamma^\mathrm{lep}_\mathrm{W,Z}$ can be accurately measured and computed, $\Gamma^\mathrm{tot}_\mathrm{W,Z}$ is often directly used to extract $\alphasmZ$. 
\item The ratio of W, Z hadronic-to-leptonic widths, defined theoretically as
\begin{eqnarray}
\mathrm{R}_\mathrm{W,Z}(Q) = \frac{\Gamma^\mathrm{had}_\mathrm{W,Z}(Q) }{\Gamma^\mathrm{lep}_\mathrm{W,Z}(Q) } =  
\mathrm{R}_\mathrm{W,Z}^\mathrm{EW} \left(1 + \sum^{4}_{i=1} a_i(Q)\left(\frac{\alphas(Q)}{\pi}\right)^i\!\!+\mathcal{O}(\alphas^5)+\delta_\mathrm{mix}
+\delta_\mathrm{np}\right)\!,
\label{eq:R_alphas}
\end{eqnarray}
where the $\mathrm{R}_\mathrm{W,Z}^\mathrm{EW}=f(\alpha,\alpha^2,\dots)$ prefactor, that depends on the fine structure constant $\alpha$, now accounts for the purely electroweak dependence of the calculation. 
Experimentally, in the W boson case the denominator of the R$_\mathrm{W}$ ratio represents the {\it sum} of all leptonic decays, and R$_\mathrm{W}$ can be accurately determined from the ratio of hadronic over leptonic decay branching ratios: R$_\mathrm{W}=\mathcal{B}_\mathrm{W}^\mathrm{had}/\mathcal{B}_\mathrm{W}^\mathrm{lep} = 2.069 \pm 0.019$~\cite{PDG}. 
However, in the Z boson case the denominator of R$_\mathrm{Z}$ is the average width over the three charged lepton species,
\ie\ R$_\mathrm{Z} = \Gamma_\mathrm{Z}^\mathrm{had}/\Gamma_\mathrm{Z}^\mathrm{\ell} = 20.767 \pm 0.025$~\cite{PDG} with $\Gamma_\mathrm{Z}^\mathrm{\ell} = \frac{1}{3}(\Gamma_\mathrm{Z}^\mathrm{e} + \Gamma_\mathrm{Z}^\mathrm{\mu} + \Gamma_\mathrm{Z}^\mathrm{\tau})$, which can be more precisely measured\footnote{This ratio is often labeled as $R^0_\ell$ but, for simplicity, we keep the R$_\mathrm{Z}$ symbol throughout the paper.}.

\item In the Z boson case, the hadronic cross section at the resonance peak in $\epem$ collisions, theoretically given by
\begin{equation}
\so = \frac{12 \pi}{m_\mathrm{Z}} \cdot \frac{\Gamma_\mathrm{Z}^{e}\Gamma_\mathrm{Z}^\mathrm{had}}{(\Gamma^\mathrm{tot}_\mathrm{Z})^2}\,,
\label{eq:sigma0Z}
\end{equation}
 where $\Gamma_\mathrm{Z}^{e}$ is its electronic width, is also used, since $\so$
 can be measured  with small experimental uncertainties in the $\epem \to \mathrm{Z} \to \mathrm{hadrons}$ process, independently of each $\Gamma^\mathrm{e,had,tot}$ width appearing in the theoretical equation.
\end{itemize}

In the expressions (\ref{eq:Gamma_alphas}) and (\ref{eq:R_alphas}), $Q=m_\mathrm{W},m_\mathrm{Z}$ is the relevant energy scale of the decay process, $a_i$ are coefficients of the pQCD expansion calculated today up to order $i = 4$ (\ie\ next-to-next-to-next-to-leading order, or N$^3$LO, accuracy), the $\mathcal{O}(\alphas^5)$ term indicates sub-permil corrections (of N$^4$LO accuracy) not yet computed, 
$\delta_{_\mathrm{EW}}=f(\alpha,\alpha^2,\dots)$ and $\delta_\mathrm{mix}=f(\alpha\alphas,\alpha\alphas^2,\alpha^2\alphas,\dots)$ correspond respectively to high-order electroweak and mixed QCD$\oplus$EW corrections, and $\delta_\mathrm{np}(\lqcd^p/Q^p)$ are power-suppressed ($p=4$) non-perturbative QCD corrections.\\

Since the Born level term in the calculation of W and Z hadronic decays is solely determined by EW parameters, all the sensitivity on $\alphas$ comes only through the small higher-order pQCD corrections. For example, for $\alphasmZ = 0.118$, the size of the pQCD sum in Eq.~(\ref{eq:R_alphas}) amounts to a $\sim$3\% effect in the calculation of $\rm R_\mathrm{W,Z}$, and thereby below permil experimental accuracies in this ratio are required for a competitive (percent level, today) $\alphasmZ$ determination~\cite{Salam:2017qdl}. Such an experimental precision has been achieved in Z boson measurements~\cite{ALEPH:2005ab}, but not in the W boson case, and that is why the latter does not yet provide a precise $\alphasmZ$ extraction~\cite{dEnterria:2016rbf} as discussed below. Reaching permil uncertainties in $\alphasmZ$ determinations requires many orders of magnitude smaller uncertainties in the experimental W and Z measurements than available today, a situation only reachable at a future high-luminosity $\epem$ collider, such as the FCC-ee, operating at the Z pole and WW threshold energies with very large integrated luminosities~\cite{FCCee}.\\

The purpose of this work is twofold. First, to implement the latest developments in the theoretical calculations of Eqs.~(\ref{eq:Gamma_alphas}) and (\ref{eq:R_alphas}), as well as in the experimental Z boson measurements, and thereby, second, extract more accurate and precise values of $\alphasmZ$ from the corresponding data-theory comparisons. The current state-of-the-art calculations of W and Z widths include ${\cal O}(\alphas^4)$, \ie\ N$^3$LO in QCD~\cite{Baikov}, plus mixed ${\cal O}(\alpha\alphas)$ QCD$\oplus$EW~\cite{Kara:2013dua,Freitas:2014hra}, as well as (in the Z case) the full two-loop ${\cal O}(\alpha^2)$ EW~\cite{Freitas:2013dpa,Dubovyk:2018rlg} and leading fermionic three-loop  ${\cal O}(\alpha^3)$ EW~\cite{Chen:2020xzx}
corrections. The latest $\alphasmZ$ extraction from the W data~\cite{dEnterria:2016rbf} employed the NNLO QCD (plus mixed QCD$\oplus$EW) result for the total W boson width, but not the full N$^3$LO expression. The latest $\alphasmZ$ extraction from the Z data~\cite{Haller:2018nnx} lacked the recent full two-loop 
and leading fermionic three-loop 
EW corrections.
In this study, we take into account all these theoretical developments and, in addition, we use the latest experimental values of Z boson pseudo-observables recently modified to account for updated LEP luminosity corrections at and off the resonance peak~\cite{Voutsinas:2019hwu,Janot:2019oyi}, and perform also a first combined analysis of the $\GWtot$ and $\RW$ data in the W boson case. Detailed studies of the propagated experimental, theoretical, and parametric uncertainties are provided, which are particularly relevant in the context of future $\alphasmZ$ extractions expected at the FCC-ee.

\section{Theoretical setup}

In our study, we first code all the state-of-the-art analytical (unintegrated) expressions for the leptonic and hadronic W boson decay widths~\cite{Denner:1991kt,Kara:2013dua} in \mathematica\ v12.0~\cite{mathematica}, making use of the {\sc LoopTools} v2.15 library~\cite{LoopTools} to carry out the corresponding loop integration for all known higher-order corrections, and derive convenient parametrizations of all quantities for the subsequent phenomenological analysis. In all our formulas derivations, we explicitly take into account the finite lepton (except for the neutrinos) and quark masses. Second, we implement the full-N$^3$LO W boson widths parametrizations, as well as the full two-loop ${\cal O}(\alpha^2)$ and leading fermionic three-loop  ${\cal O}(\alpha^3)$ EW corrections for the Z boson, into the \gfitter\ code~v2.2~\cite{Haller:2018nnx},
which is then used to carry out the data-theory fits and corresponding extractions of $\alphasmZ$. 
For all numerical evaluations, we use the latest values of the SM parameters and their
associated uncertainties~\cite{PDG}: 
\begin{alignat*}{7}
  m_u &= 2.16^{+ 0.49}_{-0.26}\MeV \, ,  &\quad m_d &= 4.67^{+ 0.48}_{-0.17}\MeV \, , \notag \\
  m_c &= 1.27 \pm 0.02 \GeV \, ,  &\quad m_s &= 93^{+ 11}_{- 5} \MeV \, , \notag \\
  m_t &= 172.9 \pm 0.4\GeV \, ,  &\quad m_b &= 4.18^{+ 0.03}_{- 0.02} \GeV \, , \notag \\
  m_\mu &= 105.6583745 \pm 0.0000024 \MeV \, ,  &\quad m_\tau &= 1.77686 \pm 0.00012 \GeV \, , \stepcounter{equation}\tag{\theequation}\label{eq:num_input} \\
  m_\mathrm{H} &= 125.10 \pm 0.14 \GeV \, ,  &\quad m_e &= 510.99895000 \pm 0.00000015 \keV \, , \notag \\
  \MW &= 80.379 \pm 0.012 \GeV \, , &\quad \MZ &= 91.1876 \pm 0.0021 \GeV \, ,	 \notag \\
 \varDelta\alpha & = 0.05903\pm0.00010  & \quad G_F &= \left(1.1663787 \pm 0.0000006\right) \cdot 10^{-5}\GeV^{-2}\,.
\end{alignat*}
Here, $m_u$, $m_d$, and $m_s$ correspond to current-quark masses, $m_c$, $m_b$, and $m_t$ to pole masses, the Higgs boson mass $m_\mathrm{H}$ is the most recent LHC average value~\cite{Aad:2015zhl}, and
$\varDelta\alpha$ is the change in the QED coupling $\alpha(Q)$ from $Q=0$ to $Q = \MZ$ as given\footnote{The value $ \varDelta\alpha  = 0.05903\pm0.00010$ quoted in Eq.~(\ref{eq:num_input}) is obtained from $\varDelta\alpha = \varDelta\alpha_\mathrm{lep}+\varDelta\alpha_\mathrm{had}$, 
with $\varDelta\alpha_\mathrm{lep}(\MZ) = 0.0314979 \pm 0.0000002$~\cite{Sturm:2013uka} and $\varDelta \alpha_{\mathrm{had}}^{(5)}(\MZ) = 0.02753 \pm 0.00010$~\cite{Davier:2019can}, and implies $\alpha^{-1}(\MZ) = 128.947 \pm 0.013$.} 
by $\alpha(\MZ) = \alpha(0)/(1-\varDelta\alpha)$. When not left free, the QCD coupling is taken at its current world average, $\alphasmZ$~=~0.1179~$\pm$~0.0010.
The experimental values of the CKM matrix elements used are those listed in Table~\ref{tab:CKM} (center column) that approximately satisfy the unitarity condition $\sum_i V_{ij}V_{ik}^* = \delta_{jk}$ and
$\sum_j V_{ij}V_{kj}^* = \delta_{ik}$. From the experimental CKM element values today, one obtains $\sum_{u,c,d,s,b} |V_{ij}|^2 = 2.024 \pm 0.032$ (with a 1.7\% uncertainty, dominated by the $\Vcs$ value), although in various cases below we will assume exact CKM unitarity. In this latter case, we will take $\sum_{u,c,d,s,b} |V_{ij}|^2 \equiv 2$ with the individual $|V_{ij}|$ values that satisfy the condition as derived from the PDG fit~\cite{PDG} (right column of Table~\ref{tab:CKM}). 

\begin{table}[H]
\caption{Current values of the CKM matrix elements determined from experimental measurements (center column) and from a global fit where CKM unitarity is assumed (right column)~\cite{PDG}.\label{tab:CKM}}
\centering
\begin{tabular}{lcc}\hline
CKM elements   & experiment        & global fit (CKM unitarity)   \\\hline 
$|V_{ud}|$     & $0.97420 \pm 0.00021$  & $0.97446 \pm 0.00010$           \\
$|V_{us}|$     & $0.2243 \pm 00005$     & $0.22452\pm 0.00044$            \\
$|V_{cd}|$     & $0.218 \pm 0.004$      & $0.22438 \pm 0.00044$           \\
$|V_{cs}|$     & $0.997 \pm 0.017$      & $0.97359_{-0.00011}^{+0.00010}$ \\
$|V_{cb}|$     & $0.0422 \pm 0.0008$    & $0.04214 \pm 0.00076$           \\
$|V_{ub}|$     & $0.00394 \pm 0.00036$ & $0.00365 \pm 0.00012$   \\
$\sum_{u,c,d,s,b} |V_{ij}|^2$ & $2.043 \pm 0.034$   & $2 \pm 0.0004$
\\\hline
\end{tabular}
\end{table}

The theoretical W and Z pseudo-observables computed via Eqs.~(\ref{eq:Gamma_alphas})--(\ref{eq:sigma0Z}) have two types of uncertainties. The first ``parametric'' one is associated with the uncertainties of the various input parameters, listed in (\ref{eq:num_input}) and Table~\ref{tab:CKM}, in the calculations. In the case of the W boson, the most important ones are $\Vcs$, $\MW$, and $\alphasmZ$; whereas for the Z boson, they are $\MZ$, $\alphasmZ$, and $\alpha$. 
The second one, of purely theoretical origin, arises from missing (QCD and/or EW) higher-order corrections in the calculations, which are increasingly small thanks to the incorporation of the new terms discussed in this work.
All uncertainties in the W and Z pseudo-observables are estimated for each boson, and propagated into the final $\alphasmZ$ values extracted, as explained in the next two sections for each electroweak boson independently.

\section{W boson observables}

The use of Eqs.~(\ref{eq:Gamma_alphas}) or (\ref{eq:R_alphas}) to extract $\alphasmZ$ from the W boson data requires the most accurate theoretical expressions available for its hadronic and leptonic widths, with their sum providing $\Gamma^\mathrm{tot}_\mathrm{W}$. Our improvements compared to previous works are discussed next.

\paragraph{Theoretical leptonic W boson decay width.}
The expression for the decay width of the W boson into leptons is based on the work~\cite{Denner:1991kt}, and schematically consists of the following contributions:
\begin{equation}
    \Gamma_\mathrm{W}^\mathrm{lep}=\Gamma^{^\mathrm{lep,Born}}_\mathrm{W} \left(\! 1 + \delta_{_\mathrm{EW}}^\mathrm{virt}+\delta_{_\mathrm{EW}}^\mathrm{brem}-\Delta_\mathrm{rad}\right)\,,
\label{GammaLep}
\end{equation}
with the Born-level leptonic width given by 
\begin{equation}
    \Gamma^{^\mathrm{lep,Born}}_\mathrm{W}=\frac{G_F}{12\pi\sqrt{2}}
    \frac{\kappa(m_\mathrm{W}^2,m_{\nu}^2,m_{\ell}^2)}{m_\mathrm{W}} \left( 2m_\mathrm{W}^2-m_{\nu}^2-m_{\ell}^2-\frac{(m_{\nu}^2-m_{\ell}^2)^2}{m_\mathrm{W}}\right)\,,
\end{equation}
where $\kappa(x, y, z)$ is the K\"all\'en function that takes as parameters the charged lepton, neutrino, and W masses squared. 
The expression is written in the $G_F$ and $m_\mathrm{W}$ scheme (rather than in the alternative $\alpha$ and weak mixing angle scheme) that reabsorb many higher-order corrections into the definition of both variables themselves, so that the calculation has smaller parametric uncertainties. The virtual EW correction to the Born width, $\delta_{_\mathrm{EW}}^\mathrm{virt}$, contains infrared and ultraviolet divergences that are cancelled out by similar diverging terms in the real (bremsstrahlung) contribution $\delta_{_\mathrm{EW}}^\mathrm{brem}$~\cite{Bloch:1937pw}. They introduce a small (loop-induced) dependence of $\Gamma_\mathrm{W}^\mathrm{lep}$ on the Higgs and top masses. The last term, $\Delta_\mathrm{rad}$, encodes the radiative corrections to the muon decay~\cite{Sirlin:1980nh,Marciano:1980pb,Sirlin:1981yz}. The computed values of the terms entering in Eq.~(\ref{GammaLep}) are tabulated in Table~\ref{tab:Gamma_W_lept_contrib}.
The inclusion of these ${\cal O}(\alpha)$ EW corrections decreases the value of the total leptonic width by $-1.3$~MeV, \ie\ by about 0.6\%, compared to the Born width. In the tau lepton case, taking into account its mass induces an additional small ($-0.2$~MeV) shift of its width, compared to the massless case. The calculations have very small parametric, 0.10~MeV, and theoretical 0.16~MeV uncertainties (for each leptonic width). The latter are estimated from the relative size of the similar two- and three-loop ${\cal O}(\alpha^2,\alpha^3)$ EW  terms for the Z boson~\cite{Dubovyk:2018rlg,Chen:2020xzx,Ayres} (Table~\ref{tab:Zobserv}), which are missing in the W calculations and are a factor 10 smaller than the EW 1-loop correction. The corresponding experimental measurements (last column of Table~\ref{tab:Gamma_W_lept_contrib}) have today uncertainties about 30 times larger than their theoretical counterparts, a fact that cries out for large data samples from a new electron-positron collider running at the $\epem\to \mathrm{W^+W^-}$ threshold.

\begin{table}[htbp!]
\centering
\caption{Leptonic decay widths (in~MeV) of the W boson into each of the three lepton families, computed in the $G_F$ and $m_\mathrm{W}$ scheme based on Eq.~(\ref{GammaLep})~\cite{Denner:1991kt}, with parametric and theoretical uncertainties listed. The last row provides the total leptonic width obtained summing all individual decay modes.
The last column quotes the corresponding experimental values~\cite{PDG}.\label{tab:Gamma_W_lept_contrib}}
\begin{tabular}{lcccc}
\hline
leptonic & $\Gamma^{^\mathrm{lep,Born}}_\mathrm{W}$ & ${\cal O}(\alpha)$ EW corrections & $\Gamma_\mathrm{W}^\mathrm{lep}$ & $\Gamma_\mathrm{W}^\mathrm{lep,exp}$\\ 
decay & (MeV) & $\delta_{_\mathrm{EW}}^\mathrm{virt}+\delta_{_\mathrm{EW}}^\mathrm{brem}-\Delta_\mathrm{rad}$  & (MeV) & (MeV) \\ 
\hline
W$ \rightarrow e^{-} +\nu_{e}$ & 227.2 & $-0.005741$ &  $225.9 \pm 0.1_\mathrm{par} \pm 0.2_\mathrm{th}$  & $223.3 \pm 5.6$ \\
W$ \rightarrow \mu^{-} +\nu_{\mu}$ & 227.2 & $-0.005741$ &  $225.9 \pm 0.1_\mathrm{par} \pm 0.2_\mathrm{th}$ & $221.6 \pm 5.5$ \\
W$ \rightarrow \tau^{-} +\nu_{\tau}$ & 227.0 & $-0.005735$  &  $225.7 \pm 0.1_\mathrm{par} \pm 0.2_\mathrm{th} $ & $237.3 \pm 6.5$ \\\hline
W$ \rightarrow \ell^{-} +\nu_{\ell}$ & 681.4 & $-0.005750$ & $677.6 \pm 0.3_\mathrm{par} \pm 0.5_\mathrm{th}$  & $682.2 \pm 10.2$\\ \hline
\end{tabular}
\end{table}

From the loop-integrated theoretical expressions, we derive useful parametrizations of the partial leptonic widths through the following formula that depends on the  masses (in~GeV) of the W, Higgs, top, and charged lepton (numerically relevant only for the $\tau$ case) particles, as follows:
\begin{equation}
\Gamma_\mathrm{W}^\mathrm{lep} = \Gamma_0 + c_1\,\Delta_\mathrm{W} + c_4\,\Delta_\mathrm{H} + c_5\,\Delta_t + c_7\, \Delta_\tau \,,
\label{eq:Gamma_parametriz}
\end{equation}
with
\begin{eqnarray}
\Delta_\mathrm{W} = \left(\frac{\MW}{80.379}\right)^3-1,\quad \Delta_\mathrm{H}=\log\left(\frac{m_\mathrm{H}}{125.10}\right),\quad
\Delta_t = \left(\frac{m_t}{172.9}\right) -1,\quad
\Delta_\tau = \left(\frac{m_\tau}{1.777}\right) -1\,, 
\label{eq:EW_params_W}
\end{eqnarray}
with the fit parameters listed in the first row of Table~\ref{tab:W_parametrizations}, valid over 3 standard deviations ($3\sigma$) around the central value of each parameter.

\begin{table}[htpb!]
\centering
\caption{Coefficients of the parametrizations for the leptonic, hadronic, and total W boson widths based on Eq.~(\ref{eq:Gamma_parametriz_full}) with the parameters defined in Eqs.~(\ref{eq:EW_params_W}) and (\ref{eq:QCD_params_W}). The last column lists the maximum deviation of the parametrization with respect to the exact analytical expressions.\label{tab:W_parametrizations}}
\resizebox{\textwidth}{!}{
\begin{tabular}{lccccccccc}\hline
W widths (GeV) & $\Gamma_0$     & $c_1$  & $c_2$  & $c_3$  & $c_4$ & $c_5$  & $c_6$ & $c_7$ & Max dev. \\ \hline
$\Gamma_\mathrm{W}^\mathrm{lep}$ & 677.59 & 671.91 & -- & -- & 0.19618 & $-3.36063$& -- & $-0.328049$ & $<0.00006$ \\ \hline
$\Gamma_\mathrm{W}^\mathrm{had}$ (exp.\ CKM) & 1440.28 & 1446.61 &  734.557 & 53.76 & --  & -- & $-1.24411$ & -- & $<0.0002$ \\ 
$\Gamma_\mathrm{W}^\mathrm{had}$ (CKM unit.) & 1410.21  &  1409.59 & -- & 52.34 & --  & --   & $-1.15932$ & -- & $<0.0002$\\ \hline
$\Gamma_\mathrm{W}^\mathrm{tot}$ (exp.\ CKM) & 2117.87 & 2028.15 & 726.281 & 50.75 & $0.20766$  & $-3.36233$ & $-0.825621$ & $-0.329354$ & $<0.0002$ \\ 
$\Gamma_\mathrm{W}^\mathrm{tot}$ (CKM unit.)  & 2087.89 & 2076.83 & -- & 52.41 & 0.21022   & $-3.36233$ & $-1.01378$  & $-0.328023$ & $<0.0009$\\ 
\hline
\end{tabular}
}
\end{table}

\paragraph{Theoretical hadronic W boson decay width.}

The hadronic W decay width is given by Eq.~(\ref{eq:Gamma_alphas}) with the known one-loop $\cO{\alphas}$ pQCD and $\cO{\alpha}$ electroweak terms~\cite{CHENG,Denner:1990tx,KNIEHL}, and two-loop $\cO{\alphas^2}$, three-loop $\cO{\alphas^3}$~\cite{Gorishny:1990vf, Surguladze:1990tg}, and four-loop  $\cO{\alphas^4}$~\cite{BCK} pQCD corrections. The combined N$^3$LO pQCD plus mixed ${\cal O}(\alpha\alphas)$ corrections were computed in Ref.~\cite{Kara:2013dua}, with (small) finite quark mass effects then evaluated in Ref.~\cite{dEnterria:2016rbf}. 
Table~\ref{tab:W_had} lists the numerical values of the Born, QCD, EW, and mixed QCD$\oplus$EW terms, as well as the full hadronic W boson width (last column). The calculated hadronic W width is $\GWhad = 1440.3 \pm 23.9_\mathrm{par} \pm 0.2_\mathrm{th}$, with dominant parametric uncertainty from the experimental value of the $\Vcs$ quark coupling strength, whose relative uncertainty of $1.7\%$~\cite{PDG} propagates into $\pm$22.5~MeV for $\GWhad$.
\begin{table}[htbp!]
\centering
\caption{Hadronic width (in~MeV) of the W boson determined in this work using the experimental CKM matrix or imposing CKM unitarity, with the individual QCD, ${\cal O}(\alpha)$ EW, and ${\cal O}(\alpha\alphas)$ mixed corrections listed. The final column gives the $\GWhad$ theoretical prediction with parametric and theoretical uncertainties.\label{tab:W_had}} 
\resizebox{\textwidth}{!}{
\begin{tabular}{lcccccccc}\hline
W hadronic width (MeV)  &  $\Gamma^\mathrm{Born}$ &  $\Gamma^{(1)}_\mathrm{QCD}$  &  $\Gamma^{(2)}_\mathrm{QCD}$ &
$\Gamma^{(3)}_\mathrm{QCD}$  &  $\Gamma^{(4)}_\mathrm{QCD}$  & $\Gamma_\mathrm{EW}$  &  $\Gamma_\mathrm{mix}$ & $\GWhad$\\ \hline
$\rm W \rightarrow \qqbar'$ (exp. CKM) & 1392.173 & 52.345 & 2.773 & $-$0.925 & $-$0.221 & $-$5.057 & $-$0.748 & 
1440.3 $\pm$ 23.9$_\mathrm{par}$ $\pm 0.2_\mathrm{th}$\\
$\rm W \rightarrow \qqbar'$ (CKM unit.) & 1363.069 & 51.251 & 2.715 & $-$0.906 & $-$0.216 & $-$4.951 & $-$0.733 & 1410.2 $\pm~0.8_\mathrm{par} \pm 0.2_\mathrm{th}$\\ 
\hline
\end{tabular}
}
\end{table}

If one assumes CKM unitarity (or, equivalently, negligible $|V_{ij}|$ uncertainties) then we obtain $\GWhad = 1410.2 \pm 0.8_\mathrm{par} \pm 0.2_\mathrm{th}$, with the second most important source of parametric uncertainty being that from $\MW$, which propagates as $\pm$0.5~MeV into $\GWhad$. Our new $\GWhad$ theoretical result has changed by $-1.2$~($+11.7$)~MeV compared to that obtained in~\cite{dEnterria:2016rbf} (without) imposing CKM unitarity, due to the updated PDG parameters. In particular, the experimental value of $\Vcs$ has increased from $0.986 \pm 0.016$ to $0.997 \pm 0.017$, and that of $\Vcd$ has decreased from $0.225 \pm 0.008$ to $0.218 \pm 0.004$, leading to a change of $+13$~MeV in the hadronic width. Also, the incorporation of the ATLAS W mass measurement~\cite{Aaboud:2017svj} leads to a world-average value of $\MW$ (and its uncertainty) that has decreased by 6~MeV (3~MeV)~\cite{PDG} compared to the results of~\cite{dEnterria:2016rbf}, propagating into a $-1$~MeV change in the W hadronic width. The theoretical uncertainties of the calculations from missing higher-order corrections are estimated as follows. The pure QCD ones are considered to be of the same
size, $\pm0.02$~MeV, as the $\mathcal{O}(\alphas^5)$ corrections assessed for the hadronic Z boson width~\cite{BCK}. The missing mixed QCD$\oplus$EW $\mathcal{O}(\alpha\alphas^2,\,\alpha\alphas^3,\,\alpha^2\alphas)$ corrections, amount to $\sim$0.2~MeV, as derived from the $\mathcal{O}(\alpha\alphas)$ result~\cite{Kara:2013dua} multiplied by an extra $\alphas$ factor~\cite{Ayres}. 
Non-perturbative effects suppressed by $\mathcal{O}(\lqcd^4/\MW^4)$, zero quark mass approximations beyond NLO~\cite{Chetyrkin:1996hm}, and residual effects due to the dependence on the CKM matrix renormalization scheme evaluated in~\cite{Almasy:2008ep}, are smaller and neglected here.

From the analytical expression including all N$^3$LO QCD, ${\cal O}(\alpha)$ EW, and ${\cal O}(\alpha\alphas)$ mixed corrections, we obtain the following useful formula for the hadronic W width as a function of all relevant parameters:
\begin{equation}
\Gamma_\mathrm{W}^\mathrm{had} = \Gamma_0 + c_1\,\Delta_\mathrm{W} + c_2\,\Delta_\mathrm{CKM} + c_3\,\Delta_{\alphas} + c_6\,\Delta_{\alphas}^2   \,,
\label{eq:GammaW_had_parametriz}
\end{equation}
where $\Delta_\mathrm{W}$, 
has been already defined in Eq.~(\ref{eq:EW_params_W}), and
\begin{eqnarray}
\Delta_{\alphas}=\frac{\alphasmZ}{0.1179}-1, \quad
\Delta_\mathrm{CKM}=
\frac{\Vcd^2+\Vcs^2}{0.218^2+0.997^2}-1\,.
\label{eq:QCD_params_W}
\end{eqnarray}
In Eq.~(\ref{eq:GammaW_had_parametriz}), we have added a linear and quadratic dependence of $\GWhad$ on $\alphasmZ$ to better reproduce the dependence on the higher-order QCD corrections.
The impact of the CKM matrix elements on $\GWhad$ is mostly dominated by the two $\Vcd$ and $\Vcs$ elements, and therefore only these two are included in the parametrization. All fitted coefficients are listed in Table~\ref{tab:W_parametrizations} and are valid over $\pm3\sigma$ variations of the current uncertainties for each parameter.

\paragraph{Theoretical total W boson decay width, and $\RW$ ratio.}

The total W boson decay width is then determined from the sum of the hadronic and leptonic ones calculated as discussed previously, and parametrized as follows:
\begin{equation}
\Gamma_\mathrm{W}^\mathrm{tot} = \Gamma_0 + c_1\,\Delta_\mathrm{W} + c_2\,\Delta_\mathrm{CKM} + c_3\,\Delta_{\alphas} + c_4\,\Delta_\mathrm{H} + c_5\,\Delta_t +  c_6\,\Delta_{\alphas}^2 +
c_7\, \Delta_\tau\,,
\label{eq:Gamma_parametriz_full}
\end{equation}
with the parameters defined in Eqs.~(\ref{eq:EW_params_W}) and (\ref{eq:QCD_params_W}) and the fitted coefficients listed in Table~\ref{tab:W_parametrizations}. The final computed N$^3$LO values of $\Gamma_\mathrm{W}^\mathrm{tot}$ (assuming or not CKM unitarity in $\Gamma_\mathrm{W}^\mathrm{had}$) 
are listed in the third row of Table~\ref{tab:W_old_new}.
 The theoretical uncertainties of the calculations, from the sum of the missing higher-order EW, mixed, and QCD corrections, are $\pm0.7$~MeV, clearly much smaller than the parametric ones.  
The theoretical widths are compared to the corresponding experimental data (last column), as well as to the NNLO parametrization~\cite{Cho:2011rk} used by default in \gfitter~2.2 (first column), which gives a result
somewhat smaller than our prediction using the experimental CKM (1.3\% times smaller, $-26$~MeV), but virtually identical enforcing CKM unitarity (0.14\%, \ie\ about 3 MeV, larger). The default \gfitter\ code does not compute the separate leptonic and hadronic W widths, nor it has N$^3$LO pQCD corrections, and neglects the (very small) theoretical uncertainties accurately estimated here for the first time.
The last row of Table~\ref{tab:W_old_new} lists our computed values for the ratio $\RW$ of leptonic-to-hadronic widths. The theoretical value computed with the experimental CKM matrix has a $\pm1.7\%$ uncertainty (basically of parametric origin), almost twice larger than the experimental uncertainty ($\pm0.9\%$) of the ratio, whereas the $\RW$ result calculated assuming CKM unitarity has a much smaller $\pm0.1\%$ uncertainty (shared in half by theoretical and parametric sources).

\begin{table}[htpb!]
\caption{Theoretical and experimental values of the leptonic, hadronic, and total W boson widths, and ratio of hadronic-to-leptonic widths. 
The second column lists the current NNLO default result in \gfitter~\cite{Cho:2011rk}, the third and fourth columns lists the N$^3$LO results of our work (Eq.~(\ref{eq:Gamma_parametriz_full}) with the coefficients of Table~\ref{tab:W_parametrizations}) assuming or not CKM matrix unitarity, with associated parametric and theoretical uncertainties. The fourth column tabulates the corresponding experimental values~\cite{PDG}. \label{tab:W_old_new}}
\resizebox{\textwidth}{!}{
\begin{tabular}{lcccccccc}\hline
W boson        & \gfitter~2.2 (NNLO) & \multicolumn{2}{c}{this work (N$^3$LO)} & experiment \\
observables    &   & (exp. CKM) & (CKM unit.) &  \\ \hline
$\Gamma_\mathrm{W}^\mathrm{lep}$ (MeV) & -- & \multicolumn{2}{c}{$677.6  \pm 0.3_\mathrm{par} \pm 0.5_\mathrm{th}$}  & $682.2 \pm 10.1$\\
$\Gamma_\mathrm{W}^\mathrm{had}$ (MeV)& --  & $1440.3 \pm 23.9_\mathrm{par} \pm 0.2_\mathrm{th}$ & $1410.2 \pm 0.8_\mathrm{par} \pm 0.2_\mathrm{th}$   & $1405 \pm 29$ \\
$\Gamma_\mathrm{W}^\mathrm{tot}$ (MeV)& $2091.8 \pm 1.0_\mathrm{par}$  & $2117.9 \pm 23.9_\mathrm{par} \pm 0.7_\mathrm{th}$ & $2087.9 \pm 1.0_\mathrm{par} \pm 0.7_\mathrm{th}$     & $2085 \pm 42$ \\
R$_\mathrm{W}$ & --  & $2.1256 \pm 0.0353_\mathrm{par} \pm 0.0008_\mathrm{th}$ & $2.0812 \pm 0.0007_\mathrm{par} \pm 0.0008_\mathrm{th}$  & $2.069 \pm 0.019$ \\ \hline
\end{tabular}
}
\end{table}

The theoretical results listed in Table~\ref{tab:W_old_new} indicate that a significant reduction of the parametric uncertainties in the calculations of $\Gamma_\mathrm{W}^\mathrm{had}$ and R$_\mathrm{W}$, in particular through new precise measurements of $\Vcs$, is as urgent as having both W boson observables with improved experimental precision.

\section{Extraction of \texorpdfstring{$\alphasmZ$}{alphaS} from W boson observables}

\begin{figure}[htbp!]
\centering
\includegraphics[width=0.49\columnwidth]{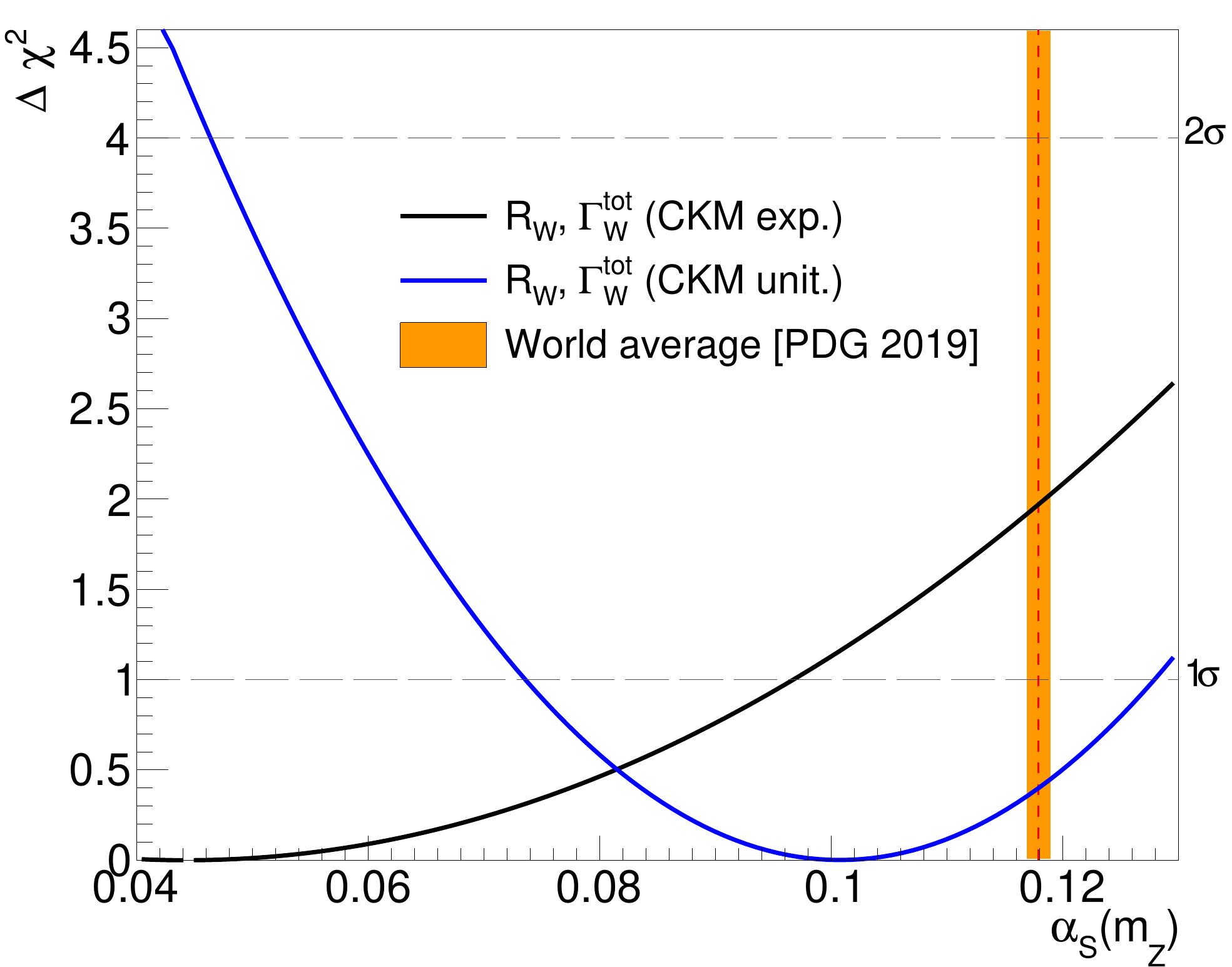}
\includegraphics[width=0.49\columnwidth]{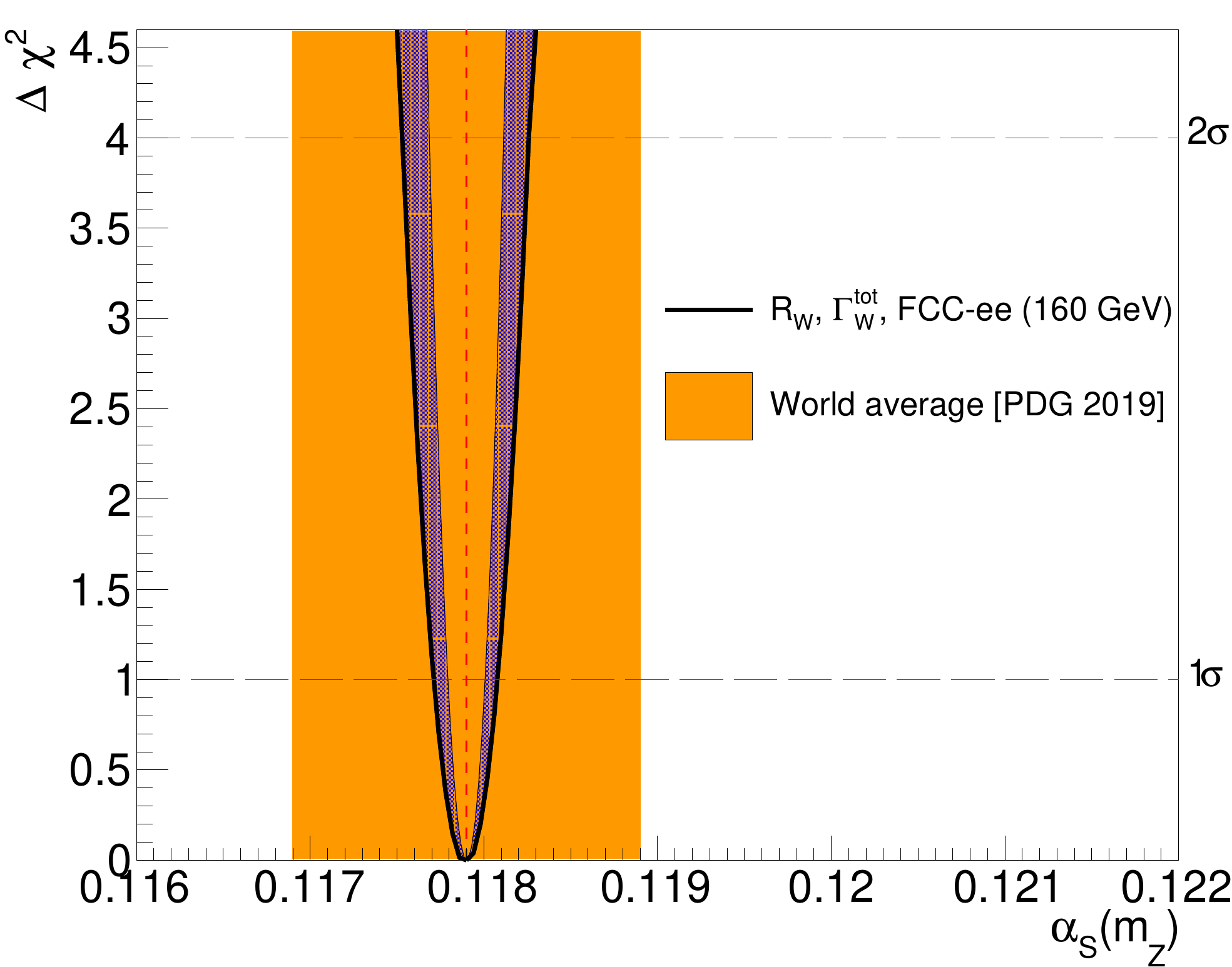}
\caption{$\Delta\chi^2$ fit profiles of the $\alphasmZ$ extracted from the combined N$^3$LO analysis of the total W width ($\Gamma_\mathrm{W}^\mathrm{tot}$) and hadronic-to-leptonic W decay ratio ($\RW$), compared to the current $\alphasmZ$ world average (vertical orange band). Left: Extraction with the present W data assuming (blue curve) or not (black curve) CKM unitarity. Right: Extraction expected at the FCC-ee, with the total (experimental, parametric, and theoretical in quadrature) uncertainties (outer parabola) and with the experimental uncertainties alone (inner parabola).
\label{fig:alphas_W}}
\end{figure}

We implement the parametrizations Eq.~(\ref{eq:GammaW_had_parametriz}) and (\ref{eq:Gamma_parametriz_full}) with the coefficients of Table~\ref{tab:W_parametrizations} into the \gfitter~2.2 program in order to extract $\alphasmZ$ from the two measured W boson observables, $\Gamma_\mathrm{W}^\mathrm{tot} = 2085 \pm 42$ and R$_\mathrm{W} = 2.069 \pm 0.019$ (combining the three leptonic decays, assuming lepton universality\footnote{Without the lepton universality assumption, this ratio is 0.6\% times smaller: R$_\mathrm{W}=2.056 \pm 0.019$.}), listed in the last column of Table~\ref{tab:W_old_new}.
Their experimental uncertainties are $2\%$ and $0.9\%$ respectively, and
combining both observables in the fit provides some improvement in the final $\alphasmZ$ precision. The results from the $\alphasmZ$ extractions are tabulated in Table~\ref{tab:alphas_W}, and the corresponding goodness-of-fit $\Delta\chi^2$ scans are plotted in Fig.~\ref{fig:alphas_W} (left). Without imposing CKM unitarity, the fitted QCD coupling constant is left basically unconstrained: $\alphasmZ = 0.044 \pm 0.052$, due to the large dominant parametric uncertainties of the theoretical $\Gamma_\mathrm{W}^\mathrm{tot}$ and R$_\mathrm{W}$ calculations (Table~\ref{tab:W_old_new}). Imposing unitary of the CKM matrix, or equivalently reducing the experimental $\Vcs$ and $\Vcd$ uncertainties to the level listed in the right column of Table~\ref{tab:CKM}, leads to an extraction with $\sim$27\% uncertainty of experimental origin. The obtained value of $\alphasmZ = 0.101 \pm 0.027$ 
(with comparatively negligible parametric and theoretical uncertainties) is, of course, in perfect accord with the current world average (orange band in Fig.~\ref{fig:alphas_W}) within the large uncertainties. With respect to the previous NNLO extraction, 
$\alphasmZ = 0.117 \pm 0.042$ based on R$_\mathrm{W}$ alone~\cite{dEnterria:2016rbf}, our new calculation leads to a $\sim$25\% relative improvement in the experimental (as well as more accurate N$^3$LO theoretical and parametric) uncertainties. 

\begin{table}[htpb!]
\centering
\caption{Values of $\alphasmZ$ extracted from the combined $\Gamma_\mathrm{W}^\mathrm{tot}$ and R$_\mathrm{W}$ measurements compared to the corresponding N$^3$LO theoretical calculations discussed in this work, assuming or not CKM unitarity, with the breakdown of propagated experimental, parametric, and theoretical uncertainties. The last row lists the $\alphasmZ$ result expected in $\epem$ collisions at the FCC-ee (see details in the text).\label{tab:alphas_W}} 
\centering
\begin{tabular}{lccccccc}\hline
W boson         & $\alphasmZ$   &   \multicolumn{3}{c}{uncertainties}     \\
observables     & extraction   & exp.  & param. &  theor.       \\\hline
$\Gamma_\mathrm{W}^\mathrm{tot}$, R$_\mathrm{W}$ (exp. CKM) & $0.044 \pm 0.052$ &$\pm 0.024$ & $\pm  0.047$ & $(\pm 0.0014)$  \\
$\Gamma_\mathrm{W}^\mathrm{tot}$, R$_\mathrm{W}$ (CKM unit.) & $0.101 \pm 0.027$ &$\pm 0.027$ & $(\pm 0.0002)$ & $(\pm 0.0016)$  \\\hline
$\Gamma_\mathrm{W}^\mathrm{tot}$, R$_\mathrm{W}$ (FCC-ee, CKM unit.) & $0.11790 \pm 0.00023$ & $\pm 0.00012$ & $\pm 0.00004$ & $\pm 0.00019$ \\
\hline
\end{tabular}
\end{table}

Achieving a truly competitive $\alphasmZ$ extraction from the W decay data, with propagated experimental uncertainties reduced by a factor of $\times$30 at least (\ie\ below the 1\% level), requires much larger data samples than those collected in $\epem$ collisions at LEP-2. At the FCC-ee, the total W width can be accurately measured through a threshold $\epem\to\rm W^{+}W^{-}$ scan around $\sqrts = 2\MW$~\cite{FCCee} center-of-mass energies, and the $\RW$ ratio will benefit from the huge sample of $5 \cdot 10^8$ W bosons (about 2.000 times larger than those collected at LEP). 
This will lead to a reduction of the $\GWtot$ and $\RW$ statistical uncertainty by more than a factor of 30, and thereby bring the propagated experimental uncertainty of $\alphasmZ$ to the desired level. Without parallel progress in the measurements of $\Vcs$, $\Vcd$, and $\MW$, the parametric uncertainty would then largely dominate the precision of any $\alphasmZ$ extraction, as it is the case today when CKM unitarity is not enforced. However, both CKM elements will be also accurately determined at the FCC-ee, exploiting the huge and ``clean'' samples of charmed mesons, and an experimental precision of 0.5 (1.2)~MeV for the W mass (width) is within reach at the FCC-ee, with 12~ab$^{-1}$ accumulated at the W pair production threshold~\cite{FCCee}.
A combined fit with our N$^3$LO \gfitter\ setup, with the following experimental values of the W observables and all other relevant parameters expected at the FCC-ee: (i) $\Gamma^\mathrm{tot}_\mathrm{W} = 2088.0 \pm 1.2$~MeV (to be compared to $2085 \pm 42$ today) and (ii) R$_\mathrm{W} = 2.08000 \pm 0.00008$ (instead of the current $2.069 \pm 0.0019$ value), (iii) CKM unitarity (or, equivalently, $\Vcs$ and $\Vcd$  uncertainties at the level of those listed in the right column of Table~\ref{tab:CKM}), and (iv) a W mass with $\MW = 80.3800 \pm 0.0005$~GeV precision, leads to $\sim$0.1\% uncertainties in $\alphasmZ$ (last row of Table~\ref{tab:alphas_W}). At such high level of experimental and parametric precision, the present propagated theoretical uncertainties would be about ten times larger than the former, although theory improvements are also expected in the coming years~\cite{Blondel:2019vdq,Blondel:2019qlh}. As per the discussion of Table~\ref{tab:W_old_new}, the theoretical effort should be aimed at computing the missing two- and three-loop ${\cal O}(\alpha^2,\alpha^3)$ EW, N$^4$LO QCD, as well as the mixed QCD$\oplus$EW $\mathcal{O}(\alpha\alphas^2,\,\alpha\alphas^3,\,\alpha^2\alphas)$ corrections, which are all of about the same size and yield today a relative theoretical uncertainty of $\sim$3--4$\cdot10^{-4}$ in the W boson observables. With a factor of 10 reduction of the theory uncertainties, a final QCD coupling extraction at the FCC-ee with a 2-permil total uncertainty is possible: $\alphasmZ = 0.11790 \pm 0.00012_\mathrm{exp} \pm 0.00004_\mathrm{par} \pm 0.00019_\mathrm{th}$ (Table~\ref{tab:alphas_W}, bottom row), where the central value quoted is arbitrarily set at the current world average.
Figure~\ref{fig:alphas_W} (right) shows the corresponding $\Delta\chi^2$ parabola for the $\alphasmZ$ extraction expected at the FCC-ee compared to the world average today (orange band), with the dashed band covering the range between taking into account all uncertainties (outer curve) and only experimental uncertainties (inner curve) .
The progress compared to the present situation, shown in the left plot, is remarkable.

\section{Z boson observables}

Accurate extractions of $\alphasmZ$ can be achieved by comparing different Z boson pseudo-observables precisely measured at LEP, such as $\Gamma_\mathrm{Z}^\mathrm{tot}$,  $\rm R_\mathrm{Z}$, and $\so$ to the corresponding theoretical calculations given by Eqs.~(\ref{eq:Gamma_alphas})--(\ref{eq:sigma0Z}). In this work, we implement several theoretical and experimental improvements with respect to the previous state-of-the-art results. 
The theory developments are discussed first. We start by including in all calculations of the partial and total Z widths\footnote{The results of~\cite{Dubovyk:2018rlg} include also theory updates for R$_\mathrm{b,c} = \Gamma^\mathrm{b,c}_\mathrm{Z}/\GZhad$,
but we do not use those to extract $\alphasmZ$, as 
their experimental uncertainties ($\sim$0.3\% for R$_\mathrm{b}$ and $\sim$1.7\% for R$_\mathrm{c}$) are at least a factor of 3 larger than for the other Z pseudo-observables, and thereby yield less precise QCD coupling constants.}, the new full two-loop electroweak terms, as parametrized in Ref.~\cite{Dubovyk:2018rlg}, plus the leading fermionic three-loop EW corrections of Ref.~\cite{Chen:2020xzx}. These 2-loop (plus, whenever available, leading 3-loop) EW corrections increase $\GZ$ by $+0.83$~MeV (plus $+0.33$~MeV), R$_\mathrm{Z}$ by $+0.0186$ (plus $+0.0009$), and $\so$ by 1~pb.
The exact numerical impact of these theoretical developments for all Z boson pseudo-observables can be seen by comparing the first two columns of Table~\ref{tab:Zobserv} where the N$^3$LO calculations of Refs.~\cite{Freitas:2014hra,Freitas:2013dpa}, implemented in \gfitter~2.2, are compared to our newer results\footnote{We note that whereas the default \gfitter\ calculates R$_\mathrm{Z}=\Gamma_\mathrm{Z}^\mathrm{had}/\Gamma_\mathrm{Z}^\mathrm{e}$, for massless leptons, and compares it with the corresponding experimental PDG ratio, we use here
R$_\mathrm{Z}=\Gamma_\mathrm{Z}^\mathrm{had}/(\frac{1}{3}(\Gamma_\mathrm{Z}^\mathrm{e} + \Gamma_\mathrm{Z}^\mathrm{\mu} + \Gamma_\mathrm{Z}^\mathrm{\tau}))$ provided theoretically by Ref.~\cite{Dubovyk:2018rlg} and experimentally by Refs.~\cite{Voutsinas:2019hwu,Janot:2019oyi}, assuming lepton universality.}, with the fourth column giving the corresponding percentage change. 
The total theoretical Z boson width changes by $+1$~MeV (a $+0.04$\% increase), the value of R$_\mathrm{Z}$ by $+0.017$ ($+0.08$\%), and that of $\so$ increases by 4~pb ($+0.01$\%). 
The theoretical errors from missing higher-order ${\cal O}(\alpha^3)$ EW, ${\cal O}(\alpha^5)$ QCD, and ${\cal O}(\alpha\alphas^2,\,\alpha\alphas^3,\,\alpha^2\alphas)$ mixed corrections quoted for all new values, are those estimated in Refs.~\cite{Dubovyk:2018rlg,Chen:2020xzx}. Their relative size is $\sim$1.5$\cdot10^{-4}$ for $\GZ$ and $\so$, and  $\sim$2.5$\cdot10^{-4}$ for $\Rlz$, \ie\ they are about a factor of two better than the corresponding theoretical calculations for the W boson pseudo-observables (Table~\ref{tab:W_old_new}), as expected, since the EW accuracy of the latter is only ${\cal O}(\alpha)$ today.
We provide also 
the separated propagated parametric uncertainties of the Z boson pseudo-observables, which are very similar in size to the theoretical ones.

\begin{table}[htpb!]
\centering
\caption{Theoretical and experimental values of the Z boson pseudo-observables $\Gamma_\mathrm{Z}^\mathrm{tot}$, $\rm R_\mathrm{Z}$, and $\so$ used in the $\alphasmZ$ extraction. The ``previous'' theory values in the first column are the N$^3$LO results~\cite{Freitas:2014hra,Freitas:2013dpa} implemented in \gfitter~2.2, those in the ``new'' column complement the latter with the 2-loop (and, in some cases, partial 3-loop) EW corrections~\cite{Dubovyk:2018rlg,Chen:2020xzx}, and the fourth column lists the percentage change from the default \gfitter\ prediction for each observable and its updated value. The three last columns list the corresponding experimental ``previous''~\cite{PDG} and ``new''~\cite{Voutsinas:2019hwu,Janot:2019oyi} results, as well as their associated percent change.
\label{tab:Zobserv}}
\resizebox{\textwidth}{!}{
\begin{tabular}{l|ccc|ccc}\hline
& \multicolumn{3}{c}{theory} &  \multicolumn{3}{c}{experiment}\\
& previous & new (this work) & change &  previous~\cite{PDG}  & new~\cite{Voutsinas:2019hwu,Janot:2019oyi} & change \\\hline
$\Gamma_\mathrm{Z}^\mathrm{tot}$ (MeV) & $2494.2 \pm 0.8_\mathrm{th}$ & $2495.2 \pm 0.6_\mathrm{par} \pm 0.4_\mathrm{th}$ & $+0.04\%$  & $2495.2 \pm 2.3$ & $2495.5 \pm 2.3$ & $+0.012\%$ \\ 
$\rm R_\mathrm{Z}$ 
& $20.733 \pm 0.007_\mathrm{th}$ & $20.750 \pm 0.006_\mathrm{par} \pm 0.006_\mathrm{th}$ & $+0.08\%$  & $20.767 \pm 0.025$ &  $20.7666 \pm 0.0247$ &  $-0.040\%$ \\ 
$\so$ (pb) & $41\,490 \pm 6_\mathrm{th}$ & $41\,494 \pm 5_\mathrm{par} \pm 6_\mathrm{th}$ & $+0.01\%$ & $41\,540 \pm 37$ &  $41\,480.2 \pm 32.5$ & $-0.144\%$ \\
\hline
\end{tabular}
}
\end{table}

On the experimental side, new studies~\cite{Voutsinas:2019hwu,Janot:2019oyi} have come up with updated LEP luminosity corrections at and off the resonance peak that modify the PDG results for the Z boson pseudo-observables $\Gamma_\mathrm{Z}^\mathrm{tot}$, and $\so$. The $\rm R_\mathrm{Z}$ ratios are unaffected by luminosity corrections, but an extra experimental significant digit has been added to match with the precision of the improved theoretical predictions. The ``previous'' PDG and ``new'' experimental values and their relative changes are listed in the three rightmost columns of Table~\ref{tab:Zobserv}. The change in $\Gamma_\mathrm{Z}^\mathrm{tot}$ is of $+0.012\%$. The impact on $\so$ is the largest of all pseudo-observables, with a $0.144\%$ reduction of the hadronic cross section at the Z peak that brings the data very close to the theoretical prediction now. The central $\rm R_\mathrm{Z}$ ratios have not changed, as aforementioned, but an extra precision digit is added now. 
Comparing uncertainty sources, one can see that the theoretical or parametric ones ($\sim$0.025\%) are about a factor of four smaller than the experimental ones ($\sim$0.1\%). Matching the uncertainties of the current theoretical state-of-the-art calls for higher precision measurements in $\epem$ collisions at the Z pole with, at least, 20 times larger data samples than those collected at LEP.

\section{Extraction of \texorpdfstring{$\alphasmZ$}{alphaS} from Z boson observables}

\begin{figure}[htbp!]
\centering
\includegraphics[width=0.62\columnwidth]{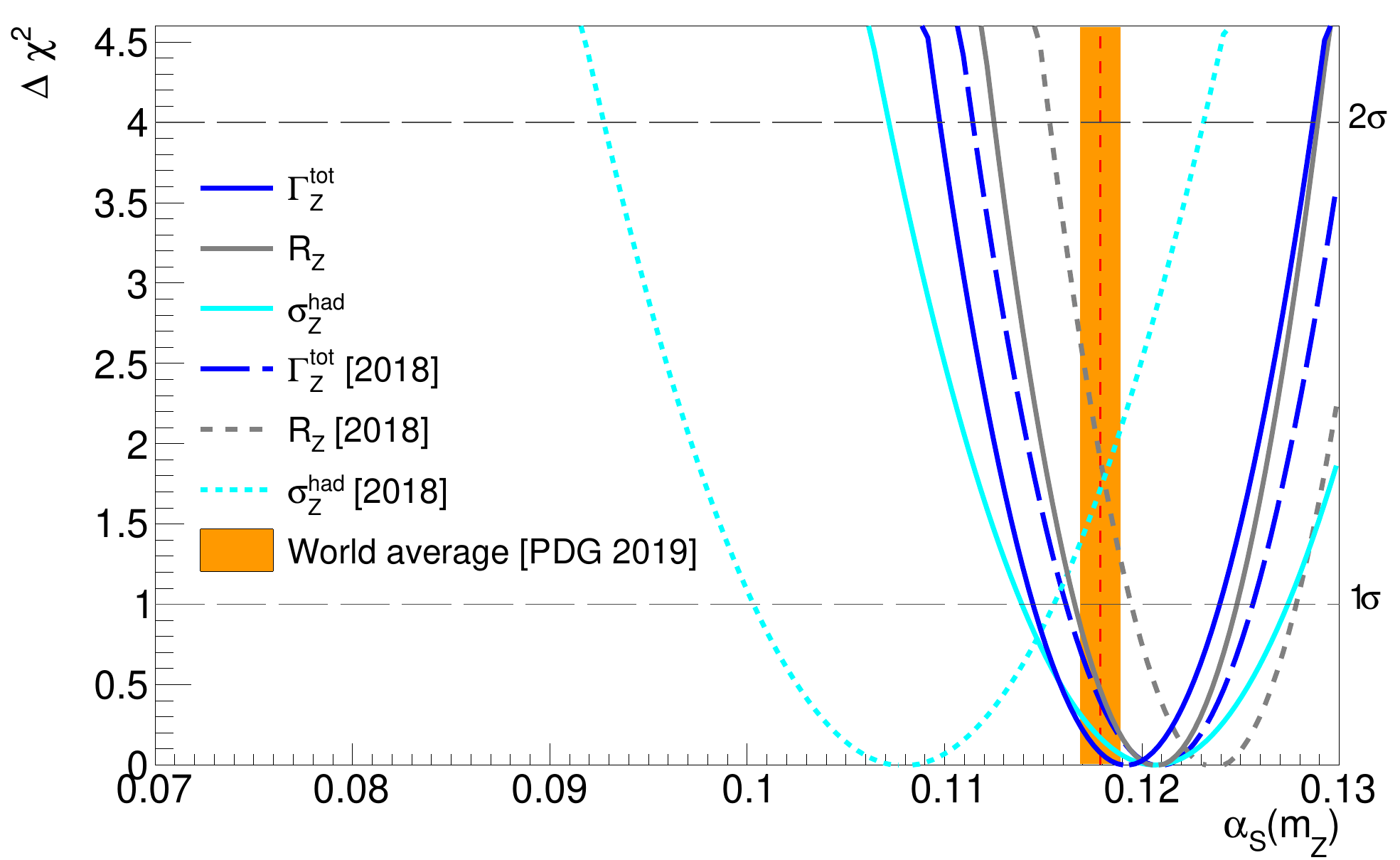}
\caption{$\Delta\chi^2$ fit profiles of $\alphasmZ$ from the N$^3$LO analysis of the Z boson pseudo-observables: $\GZ$ (blue line), $\Rlz$ (grey), $\so$ (cyan), compared to the current world average (orange band). The corresponding results obtained by \gfitter\ in 2018~\cite{Haller:2018nnx}, \ie\ without the theoretical and experimental improvements presented here, are shown in dashed lines (with the same color coding).
\label{fig:alphas_Z}}
\end{figure}

The present PDG world average~\cite{PDG} derives the $\alphasmZ = 0.1199 \pm 0.0029$ value of the so-called ``electroweak category'' by averaging the two QCD couplings obtained by letting $\alphasmZ$ as single free parameter in: (i) the 2018 global SM \gfitter\ fit~\cite{Haller:2018nnx}, and (ii) the 2019 electroweak PDG fit from all data at the Z peak~\cite{PDG}. The result from the full SM fit, $\alphasmZ = 0.1194 \pm 0.0029$ with a $\sim$2.4\% uncertainty, is slightly more imprecise than the latter, $\alphasmZ = 0.1203 \pm 0.0028$ ($\sim$2.3\% uncertainty) that is also less prone to potential biases from new physics present in other (\eg\ Higgs) sectors of the SM~\cite{alphas_confs}. 
In both cases, the measurements most sensitive to the QCD coupling constant are $\Gamma_\mathrm{Z}^\mathrm{tot}$, R$_\mathrm{Z}$, and $\so$, and the \gfitter~team has also often provided individual $\alphasmZ$ fits from each of these observables~\cite{Haller:2018nnx,Flacher:2008zq,Baak:2014ora}. In this work, we derive $\alphasmZ$ from the $\Gamma_\mathrm{Z}^\mathrm{tot}$, R$_\mathrm{Z}$, and $\so$ results, individually and combined, 
as well as from the global SM fit, introducing in all cases the complete 2-loop (and leading fermionic 3-loops for $\GZ$) EW corrections on the theory side, and using the latest experimental results listed in Table~\ref{tab:Zobserv}.
The extraction of $\alphasmZ$ is carried out with 1-D scans of this variable as a free parameter using single 
and combined observables with our updated version of \gfitter~2.2. The results from these fits are compared to the previous state-of-the-art in Table~\ref{tab:alphas_Z}, and the corresponding $\Delta\chi^2$ profiles are plotted in Fig.~\ref{fig:alphas_Z}.
The solid lines represent the results of the present improved calculations and data, whereas the dashed lines are those obtained with \gfitter\ in 2018~\cite{Haller:2018nnx}.
\begin{table}[htpb!]
\centering
\caption{Values of $\alphasmZ$ extracted at N$^3$LO accuracy from the Z boson pseudo-observables and from the global SM fit, with the default \gfitter\ 2.2 code (second column) and with the updated calculations and data considered in this work (third column). The last column lists the percent $\alphasmZ$ change of our new result compared to previous extractions~\cite{Haller:2018nnx}.\label{tab:alphas_Z}}
\begin{tabular}{lccc}\hline
Z boson        & \multicolumn{2}{c}{$\alphasmZ$ extraction}  & relative   \\
observable     & previous &  this work         & change \\ \hline
$\Gamma_\mathrm{Z}^\mathrm{tot}$ & $0.1209 \pm 0.0048$ & $0.1192 \pm 0.0047$  & $+1.4$\%  \\ 
R$_\mathrm{Z}$ & $0.1236 \pm 0.0042$ & $0.1207 \pm 0.0041$  & $-2.3$\% \\ 
$\so$ & $0.1079 \pm 0.0076$ & $0.1206 \pm 0.0067$ & $+11.8$\% \\ 
All combined & $0.1205 \pm 0.0030$ & $0.1203 \pm 0.0028$ & $-0.17$\% \\ \hline
Global SM fit & $0.1194 \pm 0.0029$ & $0.1202 \pm 0.0028 $ & $+0.58$\%\\ \hline
\end{tabular}
\end{table}
All new QCD couplings are clustered around $\alphasmZ = 0.1200$, whereas previously the extraction from $\so$ was about 2$\sigma$ lower (and also had larger uncertainties) than the average of the three, and that from $\Rlz$ was 1$\sigma$ above it. Among $\alphasmZ$ extractions, 
the most precise is that from $\Rlz$ (3.4\% uncertainty), followed by that from $\GZ$ (3.9\% uncertainty), and $\so$ (5.6\% uncertainty). The precision did not change appreciably compared to the previous $\GZ$ and $\Rlz$ results, but the extraction from the hadronic Z cross section has been improved by about 20\% thanks mostly to the updated LEP data. Table~\ref{tab:alpha_s_from_Z_detailed} lists all results with their propagated uncertainties broken down into experimental, parametric, and theoretical sources.

\begin{table}[htpb!]
\centering
\caption{Values of $\alphasmZ$ extracted at N$^3$LO accuracy from $\Gamma_\mathrm{Z}^\mathrm{tot}$, R$_\mathrm{Z}$, and $\so$ 
individually, combined, as well as from a global SM fit, with propagated experimental, parametric, and theoretical uncertainties broken down. The last two rows list the expected values at the FCC-ee from all Z pseudo-observables combined and from the corresponding SM fit (see text for details). \label{tab:alpha_s_from_Z_detailed}}
\begin{tabular}{lcccc}\hline
Z boson        & $\alphasmZ$   & \multicolumn{3}{c}{uncertainties}\\
observable     & extraction    & exp.  &  param. & theor.      \\\hline
$\Gamma_\mathrm{Z}^\mathrm{tot}$ & $0.1192 \pm 0.0047$ & $\pm0.0046$ & $\pm0.0005$ & $\pm0.0008$ \\ 
R$_\mathrm{Z}$ & $0.1207 \pm 0.0041$ & $\pm0.0041$ & $\pm0.0001$ & $\pm0.0009$ \\ 
$\so$ & $0.1206 \pm 0.0068$ & $\pm0.0067$ & $\pm0.0004$ & $\pm0.0012$ \\ 
All combined & $0.1203 \pm 0.0029$ & $\pm0.0029$ & $\pm0.0002$ & $\pm0.0008$ \\ 
Global SM fit & $0.1202 \pm 0.0028$ & $\pm0.0028$ & $\pm0.0002$ & $\pm0.0008$ \\ \hline
All combined (FCC-ee)  & $0.12030 \pm 0.00026$ & $\pm0.00013$ & $\pm0.00005$ & $\pm 0.00022$\\ 
Global SM fit (FCC-ee) & $0.12020 \pm 0.00026$ & $\pm0.00013$ & $\pm0.00005$ & $\pm 0.00022$\\\hline
\end{tabular}
\end{table}

\begin{figure}[htbp!]
\centering
\includegraphics[width=0.495\columnwidth]{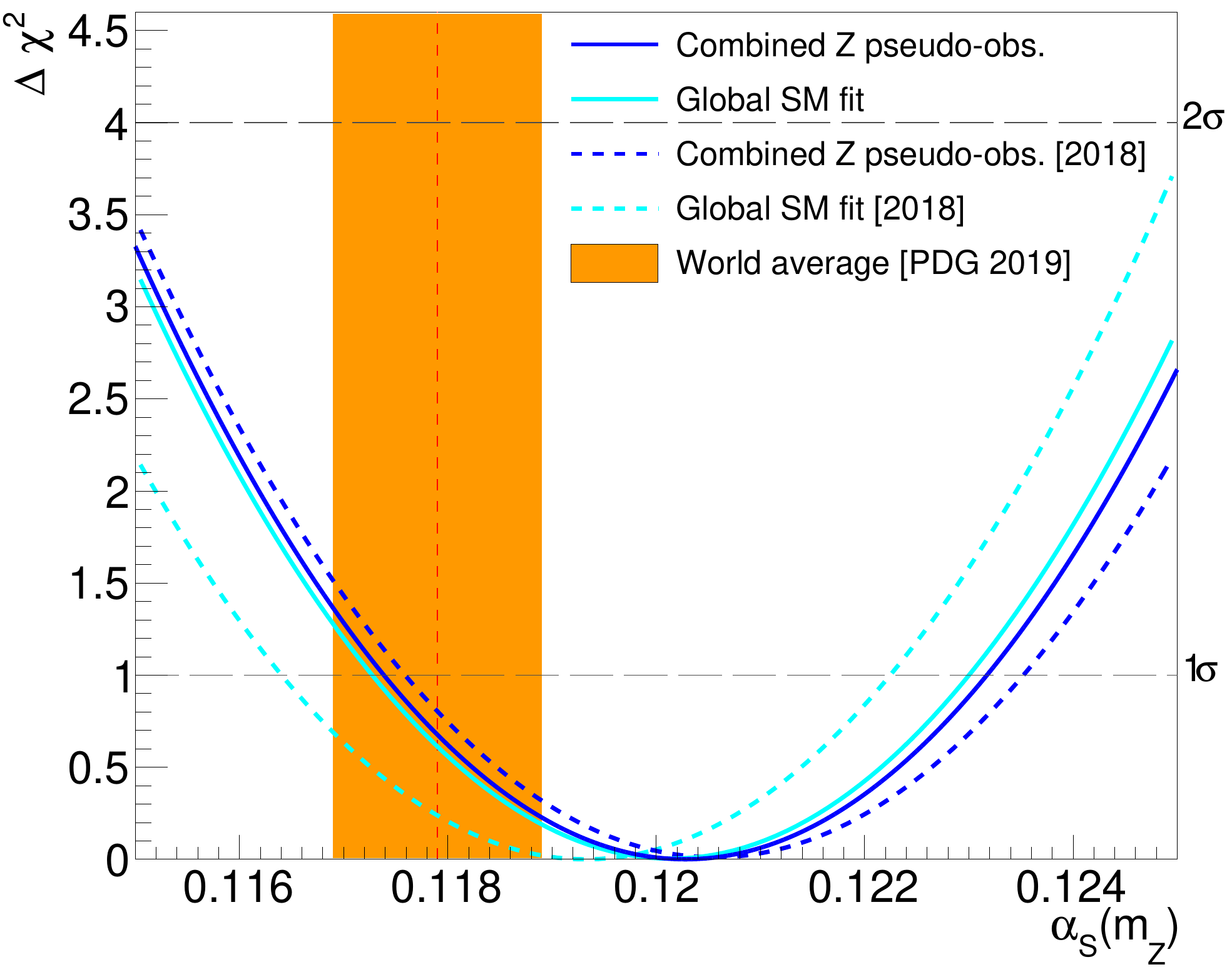}
\includegraphics[width=0.49\columnwidth]{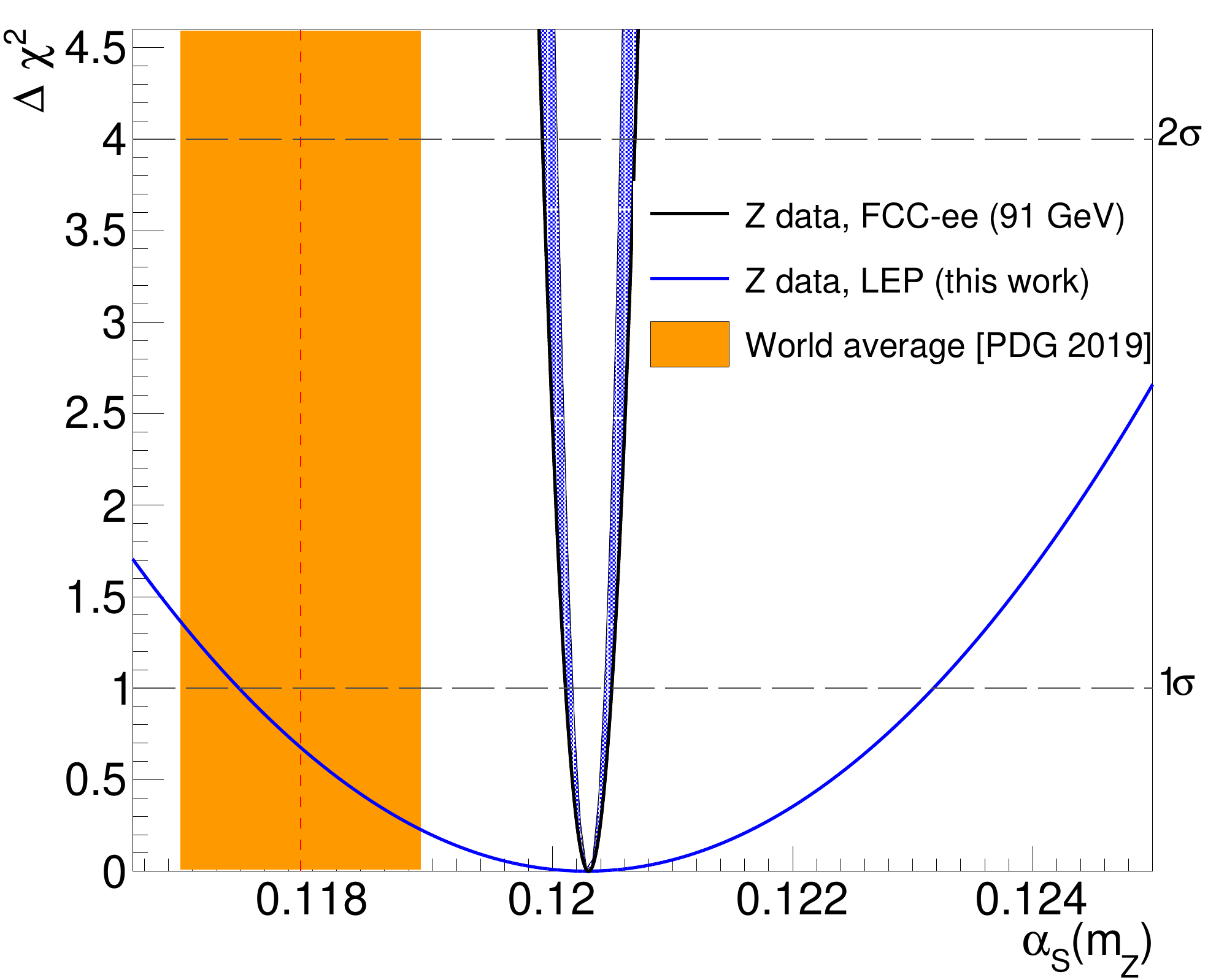}
\caption{$\Delta\chi^2$ fit profiles of $\alphasmZ$ extracted from the combined Z pseudo-observables analysis and/or the global SM fit compared to the current world average (orange band). 
Left: Current results (solid lines) compared to the previous 2018 fit (dashed lines).
Right: Extraction expected at the FCC-ee --with central value (arbitrarily) set to $\alphasmZ = 0.12030$ and total (experimental, parametric, and theoretical in quadrature) uncertainties (outer parabola) and experimental uncertainties alone (inner parabola)-- compared to the present one from the combined Z data (blue line).
\label{fig:alphas_Z_FCC}}
\end{figure}

When combining various Z observables, their associated correlation matrix is used in the fit. The LEP measurement of $\so$ has a $-32.5\%$ correlation factor with that of $\GZ$, and of $19.6\%$ with that of $\Rlz$, whereas $\GZ$ and $\Rlz$ are almost uncorrelated ($0.23\%$)~\cite{Janot:2019oyi}. 
Our extracted $\alphasmZ$ value from the combined fit of Z boson pseudo-observables is $\alphasmZ = 0.1203 \pm 0.0028$, with a $\pm$2.3\% uncertainty of almost experimental origin alone. The central value has barely changed compared to the 2018 \gfitter\ result, but the precision of our extraction has improved by about 7\% with respect to the previous state-of-the-art. Furthermore, we perform a full fit in which all SM parameters are fixed and only the QCD coupling is left free, obtaining $\alphasmZ = 0.1202 \pm 0.0028$. This result indicates a slight increase of the central $\alphasmZ$ value compared to the 2018 \gfitter\ result, with a slightly reduced final uncertainty (2.3\% compared to 2.4\% before)~\cite{Haller:2018nnx}. Both global extractions are listed in the bottom rows of Tables~\ref{tab:alphas_Z} and~\ref{tab:alpha_s_from_Z_detailed}, and the corresponding $\Delta\chi^2$ profiles are plotted in the left plot of Fig.~\ref{fig:alphas_Z_FCC} (solid lines) compared to the 2018 results (dashed lines).
Our final values, $\alphasmZ = 0.1203 \pm 0.0028$ from the combined Z boson data, and $\alphasmZ = 0.1202 \pm 0.0028$ from the full SM fit, and the PDG electroweak fit result ($\alphasmZ = 0.1203 \pm 0.0028$)~\cite{PDG} are all virtually identical.

\clearpage

At the FCC-ee, combining the $3\cdot10^{12}$ Z bosons decaying hadronically, available by integrating 100~ab$^{-1}$ at the Z pole, and the 0.1-MeV accurate $\sqrts$ calibration using resonant depolarization~\cite{Blondel:2019jmp}, will provide measurements with unparalleled precision. The statistical uncertainties in the Z mass and width, today of $\pm$1.2~MeV and $\pm$2~MeV (dominated by the LEP beam energy calibration), will be reduced to below $\pm$4~keV and $\pm$7~keV respectively, with data taken at $\sqrts = 87.9$, 91.2, and 93.9~GeV.
Similarly, the statistical uncertainty in $\RZexp$ will be negligible and the measurement in the Z\,$\to\mu^+\mu^-$ final state alone, yielding an experimental precision of 0.001 from the knowledge of the detector acceptance, will suffice to reach an absolute (relative) uncertainty of 0.001 ($5\cdot10^{-5}$) on the ratio of the hadronic-to-leptonic partial Z widths. Thus, accounting for the dominant experimental systematic uncertainties at the FCC-ee, we expect\footnote{For $\Delta \GZ$, the latest studies~\cite{Blondel:2019jmp} indicate that the ultimate systematic precision is $\sim$25~keV, \ie\ four times smaller than considered here, but the $\alphasmZ$ extraction will be anyway dominated by theoretical uncertainties.}: $\Delta \MZ$~=~0.1~MeV, $\Delta \GZ = 0.1$~MeV, $\Delta\so = 4.0$~pb, and $\Delta \Rlz  = 10^{-3}$
relative uncertainties~\cite{FCCee,Blondel:2019yqr}. In addition, the QED coupling at the Z peak will be measured with a precision of $\Delta \alpha  = 3\cdot10^{-5}$~\cite{Janot:2015gjr}, thereby also reducing the corresponding propagated parametric uncertainties. Implementing the latter uncertainties into our updated \gfitter\ setup, namely taking $\GZ = 2495.2 \pm 0.1$~MeV, $\so = 41\,494\pm 4$~pb, and $\Rlz = 20.7500 \pm 0.0010$, as well as $\MZ =  91.18760 \pm 0.00001$~GeV, and $\varDelta \alpha_{\mathrm{had}}^{(5)}(\MZ) = 0.0275300 \pm 0.0000009$, 
we derive the results listed in the last two rows of Table~\ref{tab:alpha_s_from_Z_detailed} where, the central $\alphasmZ$ value is (arbitrarily) taken at the current SM global fit extraction.
The final uncertainties in the QCD coupling constant are reduced to the $\sim$0.1\% level, namely about three times smaller than the propagated theoretical uncertainties today. Theoretical developments in the years to come should further bring down the latter by a factor of four~\cite{Blondel:2019vdq,Blondel:2019qlh}. A final QCD coupling constant extraction at the FCC-ee with a 2-permil total uncertainty is thereby reachable: $\alphasmZ = 0.12030 \pm 0.00013_\mathrm{exp} \pm 0.00005_\mathrm{par} \pm 0.00022_\mathrm{th}$ (Table~\ref{tab:alphas_Z}). Figure~\ref{fig:alphas_Z_FCC} (right) shows the $\Delta\chi^2$ parabola for the $\alphasmZ$ extraction from the Z boson data (or from the SM fit that is almost identical) expected at the FCC-ee (with the central value arbitrarily set to its present result), compared to the same extraction today (blue parabola) and to the world average (orange band). 
The large improvement, by more than a factor of ten, in the FCC-ee extraction of $\alphasmZ$ from the Z boson data will enable searches for small deviations from the SM predictions that could signal the presence of new physics contributions. 


\section{Summary}

Two new improved extractions of the QCD coupling constant at the Z pole, $\alphasmZ$, have been derived from detailed comparisons of updated experimental data on W and Z inclusive hadronic decays to state-of-the-art perturbative Quantum Chromodynamics (QCD) predictions at next-to-next-to-next-to-leading order (N$^{3}$LO) accuracy, incorporating the latest theoretical developments. 
For the Z boson case, we implement the full two-loop and leading fermionic three-loop electroweak (EW) corrections recently made available. For the W boson extraction, the hadronic and total boson widths computed at full N$^{3}$LO are used for the first time. As byproducts of our study, useful phenomenological parametrizations of the hadronic, leptonic, and total W boson widths, at the highest theoretical accuracy available today, are provided. Apart from incorporating the newest theoretical developments, we also use the latest experimental values of Z boson pseudo-observables recently modified to account for updated LEP luminosity corrections at and off the Z resonance peak. Detailed estimates of the propagated experimental, theoretical, and parametric uncertainties of the extracted $\alphasmZ$ values are provided, including perspectives for the Future Circular Collider (FCC) running with $\epem$ collisions at the Z peak and WW threshold. All the theoretical and experimental improvements are incorporated into an updated version of the \gfitter~2.2 code in order to perform the final $\alphasmZ$ fits.\\

From the combined W boson data, the total width $\GWtot$ and the ratio of hadronic-to-leptonic branching fractions $\RW$, a value of $\alphasmZ = 0.101 \pm 0.027$ is derived with still large (experimental) uncertainty, but reduced by about 25\% compared to previous extractions. A combined reanalysis of various Z boson pseudo-observables --total width $\GZ$, ratio of hadronic-to-leptonic widths $\Rlz$, and hadronic peak cross section $\so$-- yields $\alphasmZ = 0.1203 \pm 0.0028$, with an overall 2.3\% uncertainty, reduced by about 7\% compared to the previous state-of-the-art. Strong coupling determinations with permil experimental uncertainties will require high-statistics W and Z bosons data samples collected at future $\epem$ colliders, such as the FCC-ee, combined with parallel improvements in the associated parametric and theoretical uncertainties of the calculations. The parametric uncertainties can be reduced to the desired level with concomitant high-precision measurements of the relevant parameters (W and Z boson masses, QED coupling at the Z pole, and $\Vcs$ and $\Vcd$ CKM elements) at the FCC-ee. Factors of ten and four improvements with respect to the current state-of-the-art in the theoretical uncertainties of the calculations of the partial and total widths of the W and Z bosons will be needed, respectively, including higher-order N$^4$LO ${\cal O}(\alphas^5)$ QCD, ${\cal O}(\alpha^2,\alpha^3)$ electroweak, and mixed QCD$\oplus$EW $\mathcal{O}(\alpha\alphas^2,\,\alpha\alphas^3,\,\alpha^2\alphas)$ corrections missing today.

\paragraph{Acknowledgments.}
We thank Ansgar Denner for communications on the W leptonic width calculations; Ayres Freitas, Janusz Gluza, and Tord Riemann for useful theoretical discussions on EW corrections; Thomas Hahn for valuable feedback on the \mathematica\  {\sc Loop-tools} package; Patrick Janot for detailed information on the updated Z boson pseudo-observables LEP values and FCC-ee expectations; and Roman Kogler and Matthias Schott for advice on \gfitter.

\end{document}